\documentclass[%
 reprint,
 amsmath,amssymb,
 aps,
]{revtex4-1}

\usepackage{graphicx}
\usepackage{dcolumn}
\usepackage{bm}

\begin{document}

\preprint{APS/123-QED}

\title{Measurement of Effective {\Large $\Delta m_{31}^2$} using Baseline Differences of \\
Daya Bay, RENO and Double Chooz Reactor Neutrino Experiments}

\author{T.J.C.~Bezerra} \email{thiago@awa.tohoku.ac.jp}
\author{H.~Furuta} \email{furuta@awa.tohoku.ac.jp}%
\author{F.~Suekane} \email{suekane@awa.tohoku.ac.jp}%
\affiliation{%
 Research Center for Neutrino Science\\
 Tohoku University, Sendai, 980-8578, JAPAN
}%


\date{ July 4, 2012.   ~Version 2: November 30, 2012 }

\begin{abstract}
In 2011 and 2012, three reactor neutrino experiments, Double Chooz, Daya Bay and RENO showed positive signals of reactor neutrino disappearance and measured a mixing parameter $\sin^22\theta_{13}$ at average baselines 1.05, 1.65 and 1.44~km, respectively.
It is possible to measure effective $\Delta m_{31}^2$ 
($\Delta m^2$ defined in two flavor oscillation formula, hereafter referenced as 
$\Delta \tilde{m}_{31}^2$)
from distortion of neutrino energy spectrum ($E$ dependence of the oscillation) in those experiments.
However, since it requires a precise energy calibration, such measurements have not been reported yet. 
$\Delta \tilde{m}_{31}^2$ can also be measured from baseline ($L$) dependence of the neutrino oscillation. 
In this paper $\Delta \tilde{m}_{31}^2$ is measured from disappearance probabilities of the three reactor experiments which have different baselines, to be 
$2.99^{+1.13}_{-1.58}(^{+0.86}_{-0.88})  \times 10^{-3}$~eV$^2$, where the errors are two (one) dimensional uncertainties.
This is consistent with $\Delta \tilde{m}_{32}^2$ measured by $\nu_{\mu}$ disappearance in accelerator experiments. 
Importance of $\Delta \tilde{m}_{31}^2$ measurement and future possibilities are also discussed. 
\end{abstract}
%
\maketitle


\section{\label{sec:Introduction} Introduction}
Neutrino oscillation is, so far, the only firm phenomenon which is not accounted for by the standard model of elementary particles, which assume neutrinos as massless. 
The neutrino oscillation is, like other oscillations, such as 
$K^0 \Leftrightarrow \overline{K^0}$ , 
$B^0 \Leftrightarrow \overline{B^0}$(CP violation), 
$d \Leftrightarrow s$ (Cabbibo angle), 
$(ud) \Leftrightarrow (du)$ (isospin),
$B \Leftrightarrow W_3$ (Weinberg angle), 
$p(\uparrow) e(\downarrow) \Leftrightarrow p(\downarrow)e(\uparrow)$ in hydrogen atom
(21~cm HI line) etc,
assumed also to carry very important physics and we should be able to learn much about our world from it. \\
\indent There are six parameters in standard three flavor neutrino oscillation~\cite{PDG10}.
Three mixing angles between flavor eigenstates and mass eigenstates: $\theta_{12}$, $\theta_{13}$, $\theta_{23}$, 
one CP violating imaginary phase: $\delta$ and
2 independent squared mass differences: $\Delta m_{jk}^2 \equiv m_j^2 - m_k^2$, where $m_i$ are neutrino masses ($m_1$, $m_2$, $m_3$) of the three mass eignestates 
($\nu_1$, $\nu_2$, $\nu_3$) 
which correspond to the largest component of $(\nu_e,\nu_{\mu}, \nu_{\tau})$, respectively.
Before 2011, $\sin^22\theta_{12}$, $\sin^22\theta_{23}$, $\Delta m_{21}^2$,   
$\left| \Delta m_{32}^2 \right|$ were measured by various experiments and observations~\cite{PDG10}.
$\theta_{13}$ was known to be small, $\sin^22\theta_{13}<0.15$, from Chooz reactor neutrino experiment~\cite{CHOOZ99}. 
In order to measure $\delta$, to determine mass hierarchy and to solve $\theta_{23}$ degeneracy, $\theta_{13}$ has to be relatively large. 
Thus finite value of $\theta_{13}$ had been eagerly sought for. 
\\
\indent The years 2011 and 2012 will be regarded as an epoch making ones of neutrino experiments.  
T2K group showed 6 $\nu_{\mu} \to \nu_e$ appearance candidates over 1.5 expected backgrounds in June 2011~\cite{T2K11}. 
MINOS group showed also an indication of  $\nu_{\mu} \to \nu_e$ appearance~\cite{MINOS_th13_11}.
Double Chooz showed an indication of the reactor neutrino disappearance in November, 2011~\cite{DC11}.  
Daya Bay and RENO showed more precise disappearances on March and April 2012, respectively~\cite{DB12, RENO12}. 
In June 2012, at neutrino conference held in Kyoto, Double Chooz~\cite{DC_Nu12}, Daya Bay~\cite{DB_Nu12}, T2K~\cite{T2K_Nu12} and MINOS~\cite{MINOS_Nu12} updated their $\sin^2 2\theta_{13}$ measurements. 
All those results show relatively large $\theta_{13}$ and have opened up a brilliant path to future neutrino experiments. \\
\indent Reactor neutrino oscillation probability is expressed as follow. 
\begin{equation}
 P_R(\bar{\nu}_e \to \bar{\nu}_e) \sim
  1- \sin^22\theta_{13} \sin^2 \frac{\Delta \tilde{m}_{31}^2}{4E_{\nu}}L, 
  \label{eq:Pnue_nue}
\end{equation}
where, $E_{\nu}$ is neutrino energy ($\sim$ a few~MeV) and $L$ is baseline (1 $\sim$ 2~km).
$\Delta \tilde{m}_{31}^2$ is an "effective" squared mass difference often mentioned as $\Delta m_{31}^2$ in two flavor analyses. 
In three flavor oscillations, it is an average of $\Delta m_{31}^2$ and $\Delta m_{32}^2$, as will be described in the next section.   
In principle both $\sin^22\theta_{13}$ and $\Delta \tilde{m}_{31}^2$, can be measured from the oscillation. 
However, all the reactor neutrino experiments use $\Delta \tilde{m}_{32}^2$, which was measured by $\nu_{\mu}$ disappearance by MINOS group~\cite{MINOS_dm31_11}, as $\Delta \tilde{m}_{31}^2$ to extract $\sin^22\theta_{13}$ since the difference between them is in an order of $|\Delta m_{21}^2/\Delta m_{32}^2| \sim 3\%$ which is smaller than current accuracy of the measurements. 
 
It is important to measure $\Delta \tilde{m}_{31}^2$ independently from $\Delta \tilde{m}_{32}^2$ to check if the standard three flavor scheme is correct. 
If there is significant difference between  $\Delta \tilde{m}_{32}^2 $ and 
   $\Delta \tilde{m}_{31}^2 $, it indicates an existence of new physics.
   Moreover if they are measured with precision of $1\%$ or better, mass hierarchy can be solved and    $\cos \delta$ may be measured as described in the next section.
 
 Experimentally, $\Delta \tilde{m}_{31}^2$ can be measured  by analyzing either $E$ dependence of the oscillation or $L$ dependence of the oscillation. 
  Both methods use independent information, namely energy distortion and normalization, so that combining both analyses, accuracy of the $\Delta \tilde{m}_{31}^2$ will improve.  
 The former requires a precise energy calibration and no results have been reported yet. 
 The latter analysis can be performed by combining currently available disappearance information at different baselines.
 \\
\indent In this paper $\Delta \tilde{m}_{31}^2$ is measured using reported 
$\sin^22\theta_{13}$   and  baseline  of each reactor experiment.
The contents of this paper is based on our poster presentation in the conference of neutrino 2012~\cite{Thiago12}.
In next section, neutrino oscillation formula is described stressing on relation 
between  $\Delta \tilde{m}_{31}^2$ and $\Delta \tilde{m}_{32}^2$ and importance of $\Delta \tilde{m}_{31}^2$ measurement will be discussed. 
 In section-III, how we treat reactor neutrino will be described. 
 In section-IV, most recent Double Chooz, Daya Bay and RENO results~\cite{DC_Nu12, DB_Nu12, RENO12} are combined and  
 $\Delta \tilde{m}_{31}^2$ is extracted. 
 In section-V, a new experiment is proposed to measure $\Delta \tilde{m}_{31}^2$ more precisely by 
 using the reactor complementarity.  

\section{\label{sec:Formula} Neutrino Oscillation Formula and Effective $\Delta m^2$}

 The mixing matrix between flavor eigenstats and mass eigenstats is~\cite{PDG10},
 \begin{widetext}
  \begin{equation}
   U_{\alpha j}= 
   \begin{pmatrix}
    c_{12}c_{13}                                & s_{12}c_{13}                                  & s_{13} e^{-i\delta} \\
   -s_{12}c_{23} - c_{12}s_{23}s_{13}e^{i\delta} & c_{12}c_{23} - s_{12}s_{23}s_{13} e^{i\delta} & s_{23}c_{13} \\
    s_{12}s_{23}-c_{12}c_{23}s_{13}e^{i\delta}   & -s_{23}c_{12} - s_{12}c_{23}s_{13}e^{i\delta}  & c_{23}c_{13}
   \end{pmatrix},
   \label{eq:MNS}
  \end{equation}
 \end{widetext}
 where $\alpha$ is an index of flavor ($\alpha = e, \mu, \tau$) and $j$ is an index of mass eigenstates ($j=1, 2, 3$), 
 $c_{jk} = \cos \theta_{jk}$, $s_{jk} = \sin \theta_{jk}$, and $t_{jk} = \tan\theta_{jk} $ will be used later. 
 $\delta$ is so called CP violating imaginary phase. 
 Currently these parameters are measured as follows~\cite{Fogli_Nu12}. 
 $\theta_{12}\sim 34^{\circ},~\theta_{23}\sim 39^{\circ},~\theta_{13}\sim 9^{\circ},
 ~\Delta m_{21}^2 \sim 7.5 \times 10^{-5} \textrm{eV}^2$ and 
 $|\Delta m_{32}^2| \sim 2.4 \times 10^{-3}$eV$^2$.

Neutrino oscillation probability going to the same flavor is expressed by following formula, 
\begin{equation}
 P(\nu_{\alpha} \to \nu_{\alpha}) = 1-4 \sum_{j>k} \left| U_{\alpha j}\right|^2 \left| U_{\alpha k}\right|^2
 \sin^2 \frac{\Delta_{jk}}{2},
 \label{eq:oscillation}
\end{equation}
where $\Delta_{jk} \equiv \frac{\Delta m_{jk}^2 L}{2E}$. 
The second term in right hand side is called disappearance probability.
This oscillation formula is valid for both neutrino and antineutrino cases. 
\\
\indent Reactor neutrino experiments use $\bar{\nu}_e$ generated by $\beta$-decays of the fission elements in the reactor core. 
Energy of the neutrino is a few MeV. 
At around the first oscillation maximum of $\Delta_{32}$, survival probability of $\bar{\nu}_e$ is expressed as, 
\begin{align}
 P&(\overline{\nu}_e \to \overline{\nu}_e) = \notag \\
 &1-\sin^22\theta_{13} \left( c_{12}^2 \sin^2 \frac{\Delta_{31}}{2} +s_{12}^2 \sin^2 \frac{\Delta_{32}}{2}  \right) + O(10^{-3}).
 \label{eq:nue_oscillation}
\end{align}
On the other hand, the survival probability of high energy $\nu_{\mu}$ which is produced by accelerator is, 
\begin{align}
 P&(\nu_{\mu} \to \nu_{\mu}) = 1 - \sin^22\theta_{23} \times \notag \\
 &
 \begin{pmatrix}
 (s_{12}^2 + s_{13}t_{23}\sin2\theta_{12}\cos\delta)\sin^2 \frac{\Delta_{31}}{2}  \\
       +(c_{12}^2 - s_{13}t_{23}\sin2\theta_{12}\cos\delta)\sin^2 \frac{\Delta_{32}}{2}  
\end{pmatrix}
      + O(10^{-2}).
 \label{eq:nmu_oscillation}
\end{align}

Usually oscillation data are analyzed by assuming two flavor oscillation formula, 
\begin{equation}
 P(\nu_{\alpha} \to \nu_{\alpha}) = 1-\sin^22\theta \sin^2 \frac{\Delta \tilde{m}^2 L}{4E},
 \label{eq:2flavor_oscillation}
\end{equation}
and the measured mass square difference corresponds to a weighted mean of $\left| \Delta m_{32}^2 \right|$ and $\left| \Delta m_{31}^2 \right|$~\cite{Nunokawa05},
\begin{align}
 \Delta \tilde{m}_{31}^2 &= c_{12}^2 \left| \Delta m_{31}^2 \right| 
                            + s_{12}^2 \left| \Delta m_{32}^2 \right|,
                            \notag \\
 \Delta \tilde{m}_{32}^2 &= (s_{12}^2 +s_{13}t_{23}\sin2\theta_{12}\cos\delta) \left| \Delta m_{31}^2 \right|
  \notag \\ 
                        &~~+(c_{12}^2 - s_{13}t_{23}\sin2\theta_{12}\cos\delta)\left| \Delta m_{32}^2 \right| .
 \label{eq:wmean_Dm2}
\end{align}
They are called effective $\Delta m^2$.
Note that $\Delta \tilde{m}^2$ is not a difference of the mass square and is positive definite. 
Since there is a relation
\begin{equation}
 \Delta m_{31}^2 = \Delta m_{32}^2 + \Delta m_{21}^2  ,
 \label{eq:3sum}
\end{equation}
in the standard three flavor scheme, 
the difference of $\Delta \tilde{m}_{31}^2$ and $\Delta \tilde{m}_{32}^2$ is 
expressed as follows. 
\begin{align}
 \frac{2(\Delta \tilde{m}_{31}^2 - \Delta \tilde{m}_{32}^2)}
 {\Delta \tilde{m}_{31}^2 + \Delta \tilde{m}_{32}^2}
 \sim &  \pm (1-s_{13}t_{23}\tan2\theta_{12}\cos \delta)
 \notag \\
 \times \frac{  2\cos2\theta_{12} |\Delta m_{21}^2|}{|\Delta m_{31}^2| + |\Delta m_{32}^2|}
 &\sim \pm 0.012 \times (1 \pm 0.3),
 \label{eq:difference_Dm2}
\end{align}
where the overall sign depends on mass hierarchy. 
If $ \Delta \tilde{m}_{31}^2 > \Delta \tilde{m}_{32}^2$, it is normal hierarchy, and vise versa. 
In order to distinguish the mass hierarchy cases, it is necessary to distinguish the separation of
1.7$\sim$3.1\% depending on $\delta$.  
 $\Delta \tilde{m}_{32}^2$ has been measured with precision of $\sim$ 3.5\%~\cite{Fogli_Nu12}.
 So far there has been no reported measurement of $\Delta \tilde{m}_{31}^2$ and this paper is on the first measurement of it. 
  If difference between $\Delta \tilde{m}_{31}^2$ and $\Delta \tilde{m}_{32}^2$ is larger than 1.6\%, 
it can not be explained by the standard three flavor oscillation scheme.  
If both $\Delta \tilde{m}_{31}^2$ and $\Delta \tilde{m}_{32}^2$ are measured with accuracy 1\% or better in the future, the mass hierarchy and $\cos\delta$ can be measured. 
 
\section{\label{sec:RNO} Reactor Neutrino Oscillation}
  In nuclear reactors uranium and plutonium perform fission reaction;  after absorbing a thermal neutron they break up into two large nuclei called fission products, and two or three neutrons which sustains the chain reaction of the fission. 
The fission products are generally neutron rich nuclei and unstable. 
They perform  $\sim 6~\beta$-decays on average before becoming stable.
In each $\beta$-decay, a $\bar{\nu}_e$ is produced.
On the other hand, 200~MeV of energy is released per fission~\cite{Bugey94}, 
which means 
$\sim 6 \times 10^{20} \bar{\nu}_e$ are produced every second in a typical power reactor with 3~GW thermal energy. 
At 1~km from such reactors, $\bar{\nu}_e$ flux amounts to 
$\sim 5 \times 10^9$/s/cm$^2$.
The energy spectrum of the reactor neutrinos is a sum of the energy spectrum of neutrinos originated from the four fissile elements.
\begin{equation}
 S_{\nu}(E_{\nu}) = \sum_{i = ^{235}\textrm{U},^{238}\textrm{U}, ^{239}\textrm{Pu}, ^{241}\textrm{Pu}}
  \beta_i f_i(E_{\nu}),
 \label{eq:flux}
\end{equation}
where $f_i(E_{\nu})$ is reactor neutrino spectrum per fission from fissile element $i$ and 
$\beta_i$ is a fraction of fission rate of fissile element $i$.
There is a relation $\sum_i \beta_i =1$.
For equilibrium light water reactors, $\beta_i$ are similar and we use the values in Bugey paper~\cite{Bugey94}, namely 
$^{235}\textrm{U}: ~^{238}\textrm{U}: ~^{239}\textrm{Pu}: ~^{241}\textrm{Pu}$ = 0.538: 0.078: 0.328: 0.056.
 In this study, $f_i(E)$ is approximated as an exponential of a polynomial function which is  defined in~\cite{Mueller11},
\begin{equation}
 f_i(E_{\nu}) \propto \exp ( \sum_{j=1}^6 \alpha_j E_{\nu}^{(j-1)}  ).
 \label{eq:spectrum}
\end{equation}
 In reactor neutrino experiments, usually organic liquid scintillator is used to detect 
 $\bar{\nu}_e$. 
 It is rich in free protons and reactor $\bar{\nu}_e$ performs inverse $\beta$-decay 
 interaction with a proton. 
 \begin{equation}
  \bar{\nu}_e + p \to e^+ + n
  \label{eq:IBD_interaction}
 \end{equation}
 This is an inverse process of neutron $\beta$-decay (IBD) with very small $q^2$ and the cross section is precisely calculated from the neutron lifetime~\cite{Vogel99}. 
 In this analysis, information of absolute normalization is not necessary.
The energy dependence of the IBD cross section is, 
\begin{equation}
 \sigma_{\textrm{IBD}}(E_{\nu}) \propto (E_{\nu}[\textrm{MeV}]-1.29)
 \sqrt{E_{\nu}^2 - 2.59 E_{\nu} +1.4}.
 \label{eq:IBD_xsection}
\end{equation}
The disappearance probability, $P_{\textrm{d}}$, can be related to the oscillation parameters such as~\cite{KASKA06}, 
\begin{align}
 P_{\textrm{d}} &= \sin^22\theta \frac{\int S_{\nu}(E) 
 \sigma_{\textrm{IBD}}(E)
 \sin^2 \left( \frac{\Delta m^2 L}{4E}\right)dE}{\int S_{\nu}(E)\sigma_{\textrm{IBD}}(E) dE}
 \notag \\
 &\equiv \sin^22\theta \Lambda (\Delta m^2L).
 \label{eq:disappearance}
\end{align}
\indent Reactor measurement of $\theta_{13}$ is a pure $\sin^2 2\theta_{13}$ measurement in contrast to  accelerator based measurements which depend on unknown parameters. 
It means that by combining reactor results and accelerator results, information of such unknown parameters can be derived~\cite{Minakata03}. 
Under such motivations, several reactor-$\theta_{13}$ experiments were proposed in the past~\cite{WhitePaper04} and now Double Chooz, Daya Bay and RENO experiments have published positive results of the reactor neutrino disappearance
and measured $\sin^22\theta_{13}$.
These experiments make use of the same concept to reduce systematic uncertainties 
significantly~\cite{Kr2Det01} over the previous experiments of Chooz~\cite{CHOOZ99} and PaloVerde~\cite{PaloVerde01}.
That is, they construct far detector(s) at around oscillation maximum and measure the "oscillated" spectrum.
On the other hand, near detector(s) with same structure as the far detector is constructed at around a few hundreds of meters from their reactors to measure the neutrino spectrum before the oscillation. 
By comparing the data taken by the near and far detectors, the only effect caused by the oscillation can be derived by canceling systematic uncertainties of reactor neutrino flux and detection efficiencies. 
The flux-weighted average baselines of the far detectors, $\langle L \rangle$, for the three reactor experiments are 1.05~km for Double Chooz, 1.44~km for RENO and 1.65~km for Daya Bay, respectively.
The average baseline of RENO experiment was calculated using neutrino flux shown 
in~\cite{RENO_Nu12} and distances between the far detector and each reactor. 
Published values are used for Daya Bay and Double Chooz.

\section{\label{sec:3result} Combination of the results from the three Reactor Experiments}
 
\begin{figure}[htbp]
 \includegraphics[height=75mm]{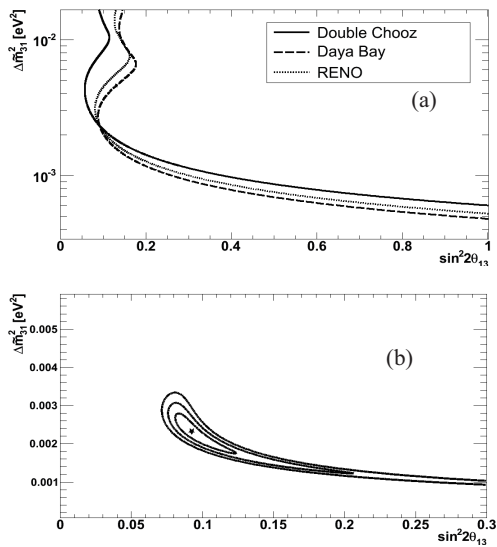}
 \caption{ Hypothetical sensitivity plots assuming true parameter values are 
 $\Delta m^2 = 2.32 \times 10^{-3}$~eV$^2$ and $\sin^2 2\theta =0.092$.
  (a) Allowed lines if disappearance probabilities are measured as expected. 
  (b) Allowed regions after combining the three reactor experiments. 
 The contour lines correspond to, from inner to outer, 1$\sigma$, 2$\sigma$ and 
 3$\sigma$ significances. 
 Disappearance error of 0.5~\% is assumed for each experiment. 
 $\Delta \tilde{m}_{31}^2$ is expected to be measured with $\sim 23(9)$~\% accuracy 
 corresponding to two (one) dimensional uncertainty.  
 }
 \label{fig:3hypo}
\end{figure}

From measured disappearance probability, an allowed line can be drawn in $\sin^22\theta-\Delta m^2$ parameter space using the relation (\ref{eq:disappearance}).
Since the baselines are different for the three reactor neutrino experiments, there are three different allowed lines as shown in fig.-\ref{fig:3hypo}(a).
The point of intersection indicates the solution of $\Delta \tilde{m}_{31}^2$ and 
$\sin^22\theta_{13}$.
In real experiments, due to errors, the three lines do not cross at same point.   
When combining different reactor results, $\chi^2$ values are calculated by using following formula for each point of the parameter space. 
\begin{equation}
 \chi^2 = \sum_{k=\textrm{exp.}} 
 \left( \frac{\sin^22\theta \Lambda(\Delta m^2 \langle L \rangle_k ) -(P_{\textrm{d}})_k}
 {\sigma_k} \right)^2,
 \label{eq:chi2}
\end{equation}
where $k$ is index of the three experiments and $\sigma_k$ is measurement error of experiment-$k$.
Fig.-\ref{fig:3hypo}(b) shows contour of the significance in case each experiment measures the disappearance with 0.5\% accuracy.
In this case, $\Delta \tilde{m}_{31}^2$ can be determined with precision of $\sim 23(9) \%$ with two (one) dimensional uncertainty.
The large difference between the one dimensional error and the two dimensional error is because the shape of one $\sigma$ contour island has long tail as shown in fig.-\ref{fig:3hypo}(b). 
Since there are two parameters to measure, at least three experiments are necessary to redundantly measure the parameters. 

In the actual analysis, $P_{\textrm{d}}$ is not directly written in papers and it is calculated from measured 
$\sin^22\theta_{13}$ and flux-weighted mean distance $\langle L \rangle$.  
In their papers, $\sin^22\theta_{13}$ were derived by assuming 
the MINOS 
$\Delta \tilde{m}_{32}^2$~\cite{MINOS_dm31_11}. 
Relations between these parameters and the disappearance probability, and allowed line are shown below.  
\begin{align}
 P_{\textrm{d}} &= \sin^22\theta_{13} \Lambda(\Delta \tilde{m}_{32}^2 \langle L \rangle_{\textrm{Far}}) 
 \notag \\
 &= \sin^22\theta \Lambda(\Delta m^2 \langle L \rangle_{\textrm{Far}}).
 \label{eq:Lambda_data}
\end{align}
 The calculated disappearance probabilities are shown in table-\ref{tab:table1} 
 together with other parameters.
\begin{table}[htbp]
 \caption{
 Parameters of the three reactor neutrino experiments.
 }
 \begin{ruledtabular}
  \begin{tabular}{lccc}
   Item              & Double Chooz & Daya Bay   & RENO \\
   \colrule
   $\langle L \rangle_{\textrm{Far}}$[km] & 1.05 & 1.65 & 1.44 \\
   $ \sin^2 2\theta_{13}  $ & $ 0.109 \pm 0.039 $ & $0.089 \pm 0.011$ & $0.113 \pm 0.023$ \\
   $P_{\textrm{d}}@ \langle L \rangle_{\textrm{Far}} $ & 
             $ 5.5\pm2.0  \%$ & $7.0 \pm 0.9 \%$  & $8.2 \pm 1.6 \% $       \end{tabular}
 \end{ruledtabular}
 \label{tab:table1}
\end{table}
 
\begin{figure}[htbp]
 \includegraphics[height=75mm]{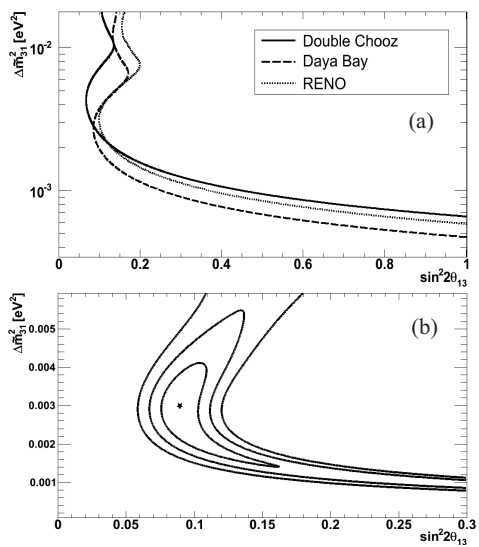}
 \caption{ 
Same as fig.-\ref{fig:3hypo} but disappearance probabilities and their errors are 
calculated from measured $\sin^22\theta_{13}$. 
 }
 \label{fig:3data}
\end{figure}

 The $\sin^22\theta_{13}$ were measured using both near and far detector at Daya Bay and RENO but only far detector was used in Double Chooz experiment. 
 It is important to point out that although the reactor experiments assume $\Delta \tilde{m}_{32}^2$ to extract $\sin^22\theta_{13}$, this analysis is independent of the assumption for the first order. 
If the experiments used different $\Delta \tilde{m}^2$, they would obtain different 
$\sin^22\theta_{13}$ but the $P_\textrm{d}$ calculated by the equation (\ref{eq:Lambda_data}) would be the same.  
$\Delta \tilde{m}_{32}^2$ was used just as a reference point. 
Fig.-\ref{fig:3data} shows the combination of the three reactor experiments calculated this way. 
The most probable oscillation parameters and their errors are, 
\begin{align}
\Delta \tilde{m}_{31}^2 &=2.99^{+1.13}_{-1.58}(^{+0.86}_{-0.88}) \times 10^{-3} \textrm{eV}^2 \notag \\
\sin^22\theta_{13} &= 0.089^{+0.071}_{-0.013}(^{+0.014}_{-0.013}),
  \label{eq: data_fixed_parameter}
\end{align}
where the errors are for two (one) dimensional uncertainty. 
This result is consistent with $\Delta \tilde{m}_{32}^2$ within one $\sigma$ and there is no
deviation from 3 flavor oscillation within this accuracy. 
The most probable $\sin^22\theta_{13}$ value coincides with the Daya Bay result
but this $\sin^22\theta_{13}$ has meaning that it was derived without assuming 
$\Delta \tilde{m}_{31}^2$. 
The minimum $\chi^2$ is 0.43 with one degree of freedom which means that the results of three reactor experiments are consistent with each other.    

\begin{figure}[htbp]
 \includegraphics[height=50mm]{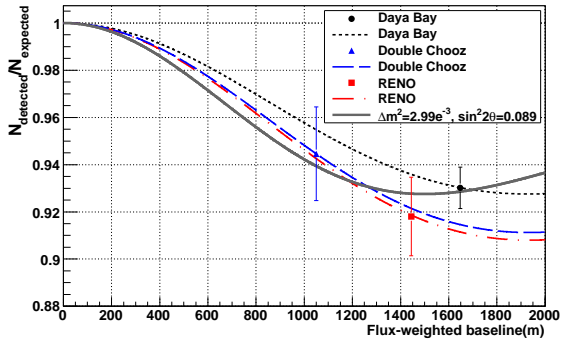}
\caption{Baseline dependence of $\bar{\nu}_e$ survival probabilities. 
Dashed and/or dotted lines are expected oscillation pattern calculated using 
$\sin^22\theta_{13}$ measured in each experiment and MINOS $\Delta \tilde{m}_{32}^2$.
The calculated disappearance probabilities correspond to the values of the expected lines at their flux-averaged baselines. 
The solid line is expectation from the most probable $\Delta m_{31}^2$ and 
$\sin^22\theta_{13}$ measured by this analysis.}
 \label{fig:baselines}
\end{figure}

Baseline dependence of observed disappearance probability and various expectation lines are  shown in fig.-\ref{fig:baselines}. 
This figure clearly shows the relation of the calculated disappearance probabilities and expected oscillation patterns.  
The meaning of disappearance probability is also described in its caption. 
In the near future, errors of the experiments are expected to improve much and the oscillation pattern will be determined much more precisely. 
%

\section{\label{sec:Future} Future possibilities}

\begin{figure}[htbp]
 \includegraphics[height=50mm]{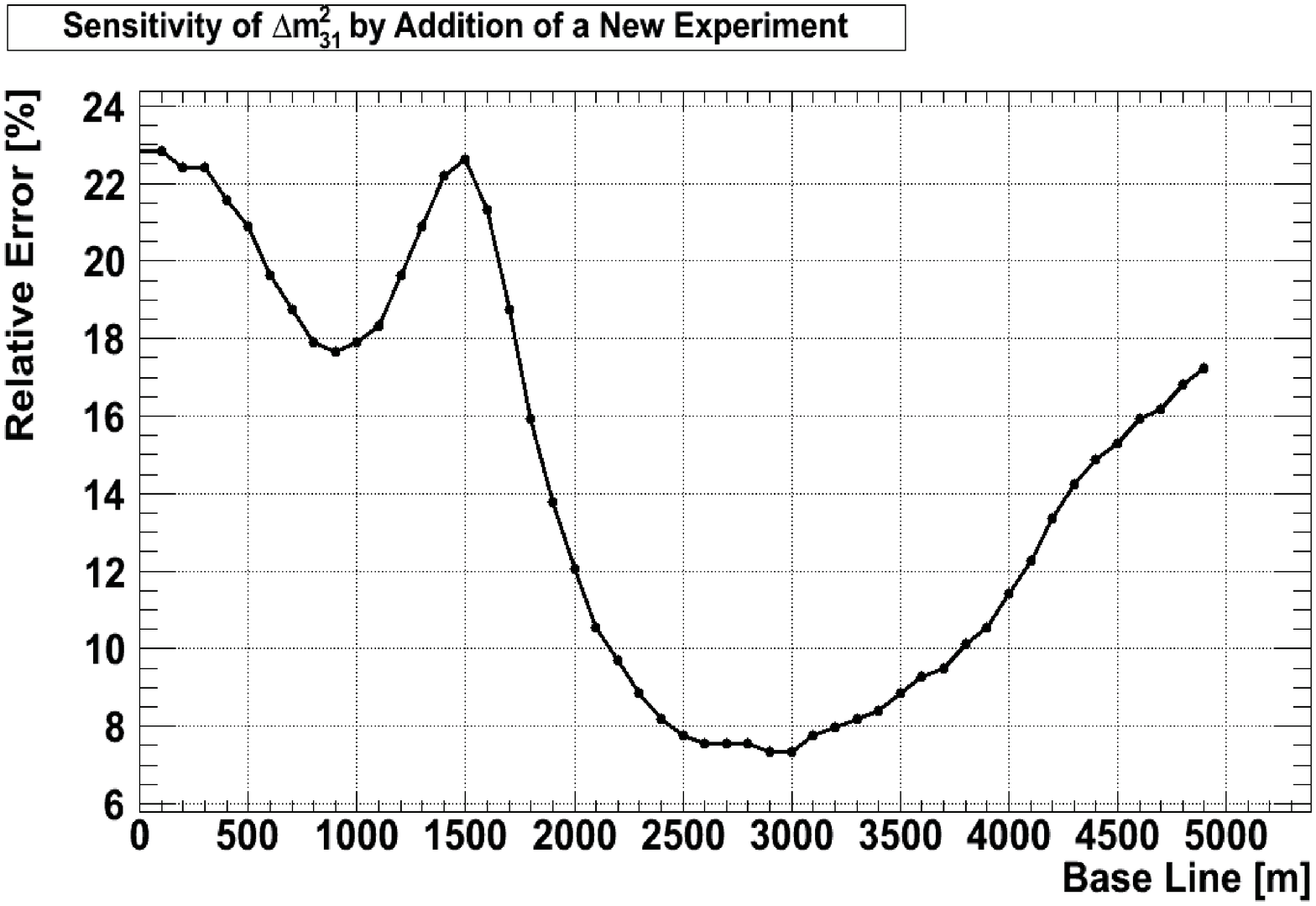}
 \caption{ Dependence of the two dimensional uncertainty on baseline of fourth experiment. }
 \label{fig:DL}
\end{figure}

\begin{figure}[htbp]
 \includegraphics[height=80mm]{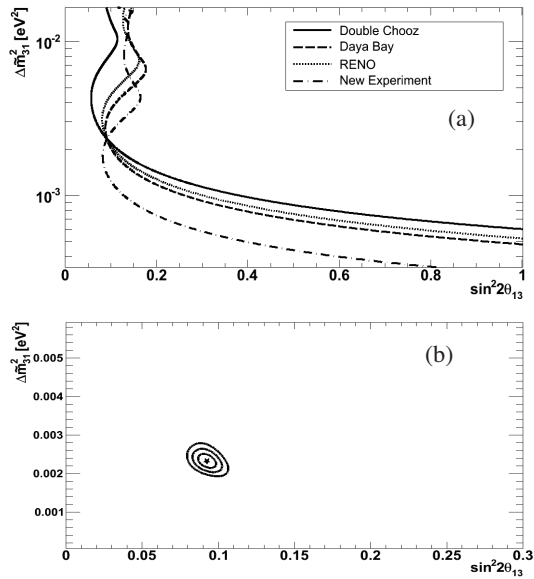}
 \caption{ Same as fig.-1 but a hypothetical fourth experiment with baseline 2.5~km is added. 
 $\Delta \tilde{m}_{31}^2$ can be measured with precision $\sim$7\%.}
 \label{fig:4hypo}
\end{figure}

It is important to evaluate how precisely we can measure $\Delta \tilde{m}_{31}^2$ since it may resolve the mass hierarchy comparing with $\Delta \tilde{m}_{32}^2$ in the future. 
In order to make the most of the reactor complementarity, we studied a case to add a fourth experiment and calculated an optimum baseline to measure 
$\Delta \tilde{m}_{31}^2 $ by combining with the current three experiments.
Fig.-\ref{fig:DL} shows dependence of the two dimensional uncertainty on baseline of the fourth experiment.
The accuracy improves rapidly when $L$ exceeds Daya Bay baselines of 1.6km and reaches to $\sim$7\% at 2.5~km. 
This is because that the tail of the island in the sensitivity contour plot vanishes
thanks to the almost perpendicular intersection of the fourth allowed line.  
Fig.-\ref{fig:4hypo} shows sensitivities with the fourth experiment with baseline 2.5~km.
Since the combined allowed region no more has tails, one and two dimensional errors become similar. 
This baseline is mere 1.5 times of Daya Bay far detector baselines and thus it is not unrealistic. 
\\
 After detailed energy calibrations are finished, the current reactor experiments will perform spectrum shape analysis to extract $\Delta \tilde{m}_{31}^2$.  
Since the shape analysis and the baseline analysis use independent information, the accuracy of the $\Delta \tilde{m}_{31}^2$ is expected to improve by combining them.
However, in order to determine the mass hierarchy, one step more improvement of the accuracy will be needed for both $\Delta \tilde{m}_{31}^2$ and $\Delta \tilde{m}_{32}^2$.

\section{\label{sec:Summary} Summary}
In this paper, measurements of $\Delta \tilde{m}_{31}^2$ by using the baseline differences between currently running reactor neutrino experiments were studied and we obtained following results. \\
\indent (1) About motivations, independent measurement of $\Delta \tilde{m}_{31}^2$ is important with following reasons. 
(i) The standard three flavor oscillation scheme can be tested.
(ii) Consistency among results from reactor neutrino experiments can be checked.
In order to perform the consistency check, at least three experiments are necessary. 
(iii) It may resolve mass hierarchy and give information of $\cos\delta$ in future experiments. 
(iv) The reactor complementarity method uses independent information from spectrum shape analysis and accuracy of $\Delta \tilde{m}_{31}^2$  will improve if they are combined.  
\\
\indent (2) The current data from Daya Bay, RENO and Double Chooz were combined and  
$\Delta \tilde{m}_{31}^2 = 2.99^{+1.13}_{-1.58}(^{+0.86}_{-0.88})$~eV$^2$ was obtained, where the first (second) error was two (one) dimensional uncertainty. 
This is consistent with $\Delta \tilde{m}_{32}^2$ measured by the accelerator experiment. 
Minimum $\chi^2$ was small which shows that the results of the three experiments were consistent with each other. 
\\
\indent (3) It was demonstrated that $\Delta \tilde{m}_{31}^2$ could be measured with precision 
$\sim$23(9)~\% by combining the three experiments if the error for their disappearance probabilities would reach 0.5\%. 
\\
\indent (4) A new detector with baseline 2.5~km will reduce both the one and two dimensional uncertainties of $\Delta \tilde{m}_{31}^2$ to $\sim 7 \%$ by combining with the current three reactor experiments.

\section*{\label{sec:Acknowledgement} Acknowledgement}
This work was supported by Ministry of Education, Culture, Sport and Technology of Japan;
Grant-in-Aid for Specially Promoted Research (20001002) and GCOE programs of Tohoku Univ.
We appreciate Dr.~H.~Nunokawa, Dr.~H.~Watanabe, Dr.~F.~Kaether,  Dr. P.~Chimenti and Double Chooz Japan group for useful discussions and comments.


\end{document}